\documentclass[prl, reprint]{revtex4-1}
\usepackage{amsmath}
\usepackage{graphicx}
\usepackage{graphicx}
\usepackage{amsmath}
\usepackage{amssymb}
\usepackage{url}
\usepackage{bm}		
\usepackage{color}	
\usepackage{textcomp}				

\begin{document}


\title{Zero-index bound states in the continuum} 



\author{Momchil Minkov}
\email[]{mminkov@stanford.edu}

\author{Ian A. D. Williamson}

\author{Meng Xiao}

\author{Shanhui Fan}
\email[]{shanhui@stanford.edu}
\affiliation{Department of Electrical Engineering, and Ginzton Laboratory, Stanford University, Stanford, CA 94305, USA}

\date{\today}

\begin{abstract}
Metamaterials with an effective zero refractive index associated to their electromagnetic response are sought for a number of applications in communications and nonlinear optics. A promising way that this can be achieved in all-dielectric photonic crystals is through the design of a Dirac cone at zero Bloch wave-vector in the photonic band structure. In the optical frequency range, the natural way to implement this design is through the use of a photonic crystal slab. In the existing implementation, however, the zero-index photonic modes also radiate strongly into the environment due to intrinsic symmetry properties. This has resulted in large losses in recent experimental realizations of this zero-index paradigm. Here, we propose a photonic crystal slab with zero-index modes which are also symmetry-protected bound states in the continuum. Our approach thus eliminates the associated radiation loss. This could enable, for the first time, large-scale integration of zero-index materials in photonic devices.
\end{abstract}

\maketitle

Recently, there has been a lot of interest in photonic metamaterials with a zero (effective) refractive index \cite{Ziolkowski2004, Alu2007, Engheta2013, Liberal2017}. In such materials, all fields oscillate in spatial unison regardless of the shape and size of the sample, and the wavelength of light appears to be effectively infinite. This has been demonstrated to lead to a variety of interesting effects and possible applications, including  tailoring the radiation phase \cite{Alu2007}, developing enhanced and directional light emitters \cite{Enoch2002, Moitra2013, Lin2016}, enhanced and directional transmission through bends, disorder, and subwavelength channels \cite{Silveirinha2006, Liu2008, Edwards2008, Edwards2009, Nguyen2010}, cloaking \cite{Hao2010}, and enhanced optical nonlinearities \cite{Suchowski2013, Alam2016, Caspani2016}. 

In the visible or near-infrared wavelength range, zero-index response can be achieved starting from the plasmonic effects from free carriers \cite{Liberal2017}, which, however also come with a substantial material loss. Alternatively, Ref. \cite{Huang2011} has shown that a zero-index effect can be achieved in a two-dimensional all-dielectric photonic crystal, whose band-structure is designed to exhibit a Dirac-like dispersion in the vicinity of the $\Gamma$-point at the Brillouin zone center. However, while a purely two-dimensional implementation of the idea of Ref. \cite{Huang2011} has intrinsically zero loss, in the optical frequency the most natural way to implement the same concept is through the use of a photonic crystal slab structure consisting of a patterned dielectric slab surrounded by low-dielectric-constant materials. In this case, the $\Gamma$-point lies in the light cone of the surrounding medium, which leads to substantial radiative losses. For example, the implementation of this concept in Ref. \cite{Vulis2017} reported a very large propagation loss of the order of 1dB/\textmu m. In Ref. \cite{Li2015}, gold mirrors on each side of the slab were introduced to avoid that radiation, but this resulted in absorption losses of comparable magnitude. Finally, it has also been proposed that this issue could be mitigated through a dielectric reflector on one side of the slab \cite{Munoz2016}, but this significantly increases the complexity and footprint of the structure. 

\begin{figure}
\centering
\includegraphics[width = \columnwidth]{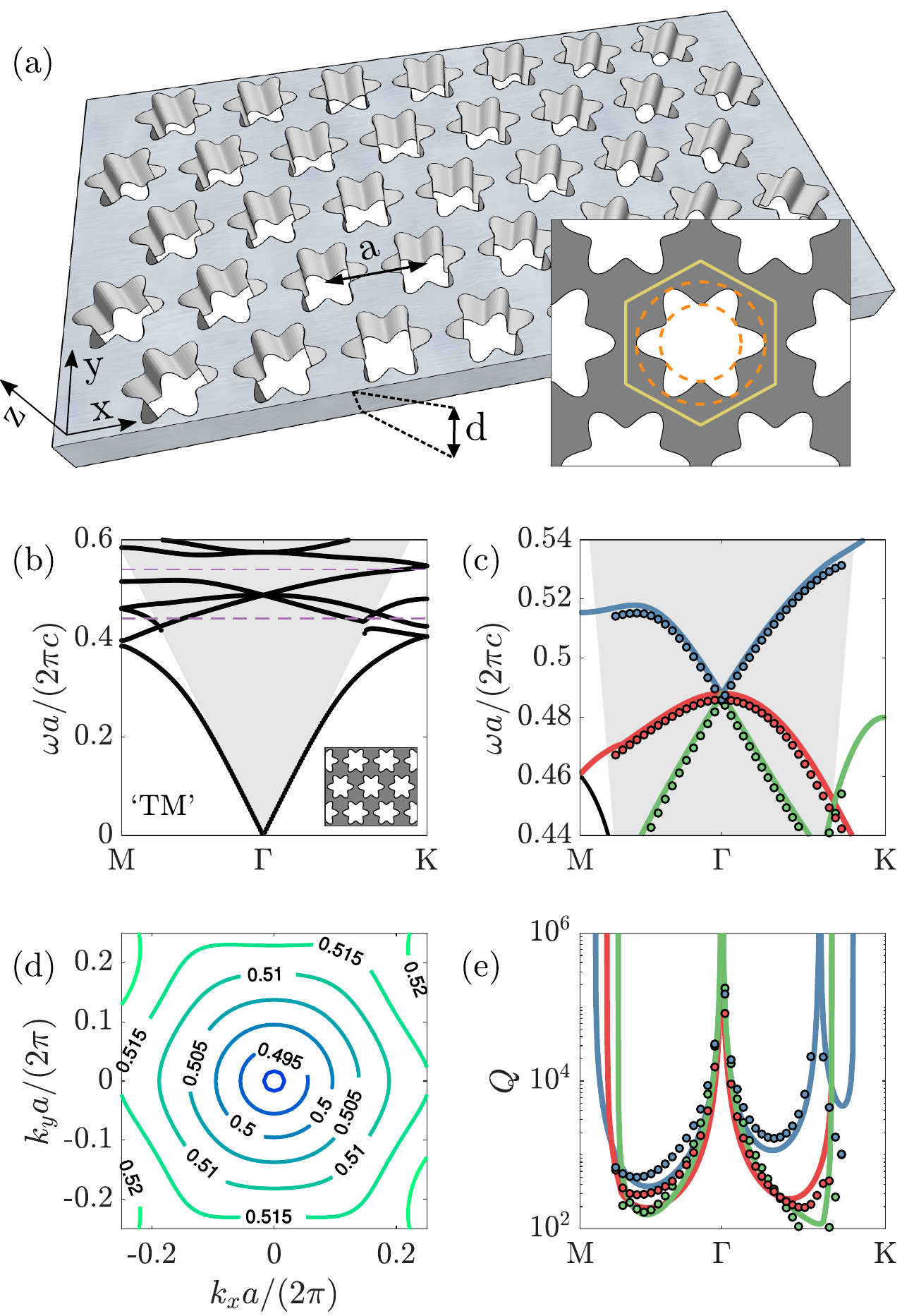}%
 \caption{(a): Schematic of the photonic crystal slab. The inset shows the `daisy' hole shape with its inner and outer radii (dashed lines), as well as the in-plane elementary cell (solid line). (b): Photonic bands for the quasi-TM modes computed with GME. (c): Zoom-in within the dashed lines of (b), with dots showing the modes computed with COMSOL. (d): Isofrequency contours in 2D reciprocal space for the top band of the dirac cone (blue band in (c)). (e): Radiative qualtiy factors for the three corresponding bands in (c).}
\label{fig:daisy}
\end{figure} 

In this Letter, we overcome the problem of radiative losses by designing a PhC slab with effectively zero-index modes, which are also symmetry-protected bound states in the continuum (BIC) \cite{Hsu2016}. Consider a photonic crystal structure with two-dimensional periodicity, with a point-group symmetry that supports at least one two-dimensional irreducible representation. At the $\Gamma$-point, suppose the crystal supports a singly degenerate state $|s\rangle$ at a frequency $\omega_1$, and a pair of doubly degenerate states $|p_x\rangle$ and $|p_y\rangle$ belonging to the two-dimensional irreducible representation at a frequency $\omega_2$. Finally, suppose also an accidental degeneracy such that $\omega_1 = \omega_2$. In the vicinity of the $\Gamma$-point, using the $k\cdot p$ method to first order, one can then show that the band-structure of the crystal is described by an effective Hamiltonian \cite{Sakoda2012}
\begin{equation}
C_{\mathbf{k}} = \begin{pmatrix}
0 & 0 & b_x k_x \\
0 & 0 & b_y k_y \\
b_x^* k_x & b_y^* k_y & 0
\end{pmatrix},
\label{eqn:Ck}
\end{equation}
where the $b_x$ and $b_y$ coefficients are given by an overlap integral between the $|s\rangle$ and $|p_{x/y}\rangle$ states and the `momentum' term, and $|b_x| = |b_y|$. Near the $\Gamma$-point the band-structure corresponding to such a Hamiltonian has two bands with Dirac-like linear dispersion, as well as a flat band. This is the same band-structure as that of a zero-index material \cite{Wang2009}. Despite the fact that the elementary cell of the photonic crystal is of comparable size to the free-space wavelength, it has been shown that an effective-medium theory can still formally be applied when the Bloch vector $\mathbf{k}$ is close to zero, i.e. when the effective wavelength is large \cite{Wu2006, Huang2011, Li2015, Vulis2017}. Therefore, an effectively zero-index material can be created using a photonic crystal structure with an appropriate accidental degeneracy at the $\Gamma$-point.

Based on the theory above, Ref. \cite{Huang2011} achieved a zero-index material starting from a photonic crystal with a square lattice as described by the $C_{4v}$ point group, which supports a single two-dimensional irreducible representation (denoted by $E$). The difficulty arises, however, when implementing this strategy in the infrared or optical wavelength range using a photonic crystal slab geometry. This is because modes at the $\Gamma$-point with a non-zero frequency lie in the radiation continuum, and, moreover modes belonging to the $E$ irreducible representation necessarily couple to external plane waves \cite{Ochiai2001, Fan2002}. As a result, the zero-index modes necessarily have significant intrinisic radiation loss, as observed experimentally in Ref. \cite{Vulis2017}.

To overcome the difficulty associated with the radiation loss, here instead we consider a hexagonal lattice with a $C_{6v}$ point group. The $C_{6v}$ group has two distinct two-dimensional irreducible representations (denoted as $E_1$ and $E_2$, respectively). For a photonic crystal slab with a $C_{6v}$ symmetry, at the $\Gamma$-point, modes that belong to the $E_2$ irreducible representation do not couple to external plane waves, i.e. these modes are symmetry-protected bound states in the continuum. Moreover, any mode at the $\Gamma$-point that belongs to a one-dimensional irreducible representation also does not couple to external plane waves. In this system, a non-radiative zero-index bound state in the continuum can be created by forcing an accidental degeneracy between modes in the $E_2$ and the $B_1$ or $B_2$ irreducible representations, since the $b_{x/y}$ coefficients in eq. (\ref{eqn:Ck}) in this case are non-zero \cite{Sakoda2012}.

We implement our proposal in the structure shown in Fig. \ref{fig:daisy}(a), which consists of a hexagonal lattice of air holes with lattice period $a$ in a silicon slab (dielectric permittivity $\varepsilon = 12$) of thickness $d = 0.5a$. The holes have a daisy-like shape that can be written in polar coordinates as $r(\phi) = r_0 + r_d\cos(6\phi)$, with $r_0 = 0.35a$ and $r_d = 0.08a$. The dashed lines in the inset show the inner and outer radii, $r_0 - r_d$, and $r_0 + r_d$, respectively. Below we refer to this structure as a ``daisy photonic crystal''. We note that a standard PhC slab with circular holes already has all the required symmetries, but we could not attain an accidental degeneracy between modes in the $E_2$ and $B_{1/2}$ representations by tuning only the circular hole radius. In contrast, we could successfully achieve such accidental degeneracy in the daisy photonic crystal by tuning $r_0$ and $r_d$. Further details regarding the way we designed this photonic crystal are provided in the Supplementary Information. 

Fig. \ref{fig:daisy}(b) shows the band-structure for the quasi-TM modes (more precisely, the anti-symmetric modes with respect to reflection from the $xy$-plane \cite{Johnson1999}) for this PhC, computed using the guided-mode expansion method (GME) \cite{Andreani2006}. Panel (c) shows a zoom-in over the region where a clear Dirac cone dispersion is visible, as well as a band that is relatively flat in the vicinity of the $\Gamma$-point. These results were also confirmed using COMSOL Multiphysics 5.3. Panel (d) shows the isofrequency contours for the high-frequency branch of the Dirac cone, and illustrates that the dispersion is highly isotropic around $\Gamma$. Finally, in panel (e) we show the quality factors (Q) associated with the radiative loss rates of the three photonic bands, computed both with GME and with COMSOL, and observe that the quality factors diverge at $\Gamma$, indicating that the states are fully bound. Note that the blue band has one additional BIC at a finite wave-vector $k$ along the $\Gamma$K direction, i.e. $k$ pointing in the $x$-direction in panel (a). We note further that the $Q$-s of all bands diverge to infinity when they enter the region outside of the light cone (compare (c) and (e)).   

\begin{figure*}
\centering
\includegraphics[width = \textwidth, trim = 0in 0in 0in 0in, clip = true]{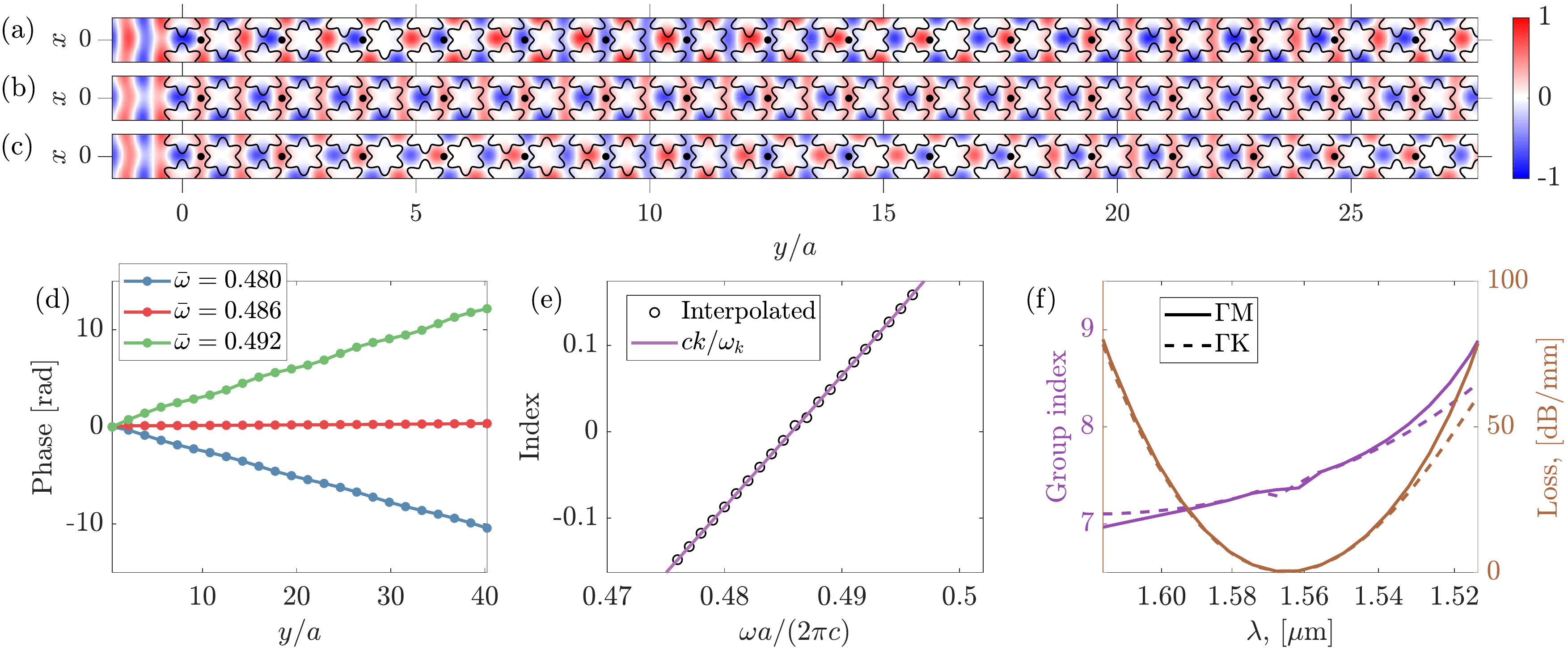}%
 \caption{(a)-(c) Electric field Re$(E_z)$ in the center of the slab ($z = 0$) for a plane-wave excitation from the left, with a wave vector along the $y$-direction ($\Gamma$M), at reduced frequency $\bar{\omega} = \omega a/(2\pi c) = $ 0.48 (a), 0.486 (b), and 0.492 (c). (d): Phase of the electric field at the sampling points marked with black dots in panels (a)-(c). (e): Effective index computed from the slope of the phase evolution with position (dots), and from the Bloch band structure (line). (f): Group index and propagation loss of the Dirac-cone bands along the two crystallographic directions as a function of wavelength, assuming $a = 760$nm. The $x$-axis scale and range in panels (e) and (f) are the same. Plots as in (a)-(e) for $\Gamma$K are in the Supplementary Information.} 
\label{fig:ind_loss}
\end{figure*}

To compute an effective index for the PhC slab, we use COMSOL to study the in-plane phase evolution of the electric field in a dielectric slab with a finite-length photonic crystal introduced in it, as shown in Fig. 2a. In Figure \ref{fig:ind_loss}(a)-(c), we show this for propagation along the $\Gamma$M direction, imposing periodic boundary conditions at $x = \pm a/2$, and perfectly-matched layer boundary conditions in the $y$- and $z$-directions. In $z$, these were placed at a distance of $3a$ from each side of the slab. The excitation frequency in panel (b), $\bar{\omega} = \omega a/(2\pi c) = 0.486$, is very close to the frequency $\omega_0$ at the tip of the Dirac cone, for which an effectively zero-index response is expected. Indeed, we see that the fields in all the elementary cells in the propagation direction oscillate in phase. In the case of panels (a) and (c), when the excitation frequency is detuned away from $\omega_0$, there is a slow evolution of the oscillation phase along the propagation direction. This is better illustrated in panel (d), where we plot the phase at equivalent reference points in each elementary cell that are related by translational symmetry (black dots in (a)-(c)). We see that the phases vary linearly as a function of propagation distance. The refractive index $n$ for a homogeneous material is such that the field evolution of a plane wave follows $E_z(\rho) = E_z(0)e^{ink_0\rho}$, where $\rho$ is the distance along the propagation direction, and $k_0 = \omega/c$. Analogously, an effective refractive index $n_e$ can be extracted from the slope of the lines in Fig. \ref{fig:ind_loss}(d), using the model $E_z(y_0 + m\sqrt{3}a) = E_z(y_0)e^{in_e k_0 m\sqrt{3}a}$, with $m$ an integer. In panel (e), we plot the effective index obtained in this way, which agrees very well the effective index $n_{eff}' = ck/\omega_k$, as obtained by directly examining the band structure as shown in Fig. \ref{fig:daisy}(b). Therefore, the bandstructure in \ref{fig:daisy}(b) explains very well the field evolution behavior, including the zero-index behavior in a finite-sized photonic crystal structure. 

In Figure \ref{fig:ind_loss}(f), we show the group index associated to the Dirac-cone bands, $n_g = v_g/c$, with the group velocity $v_g =\mathrm{d}\omega_k/\mathrm{d}k$ computed numerically. Furthermore, to show the significance of our structure for potential applications in optical information processing, we set $a = 760$nm, such that the zero-index response is in the telecommunication band around wavelength $\lambda = 1550$nm, and we compute the theoretical loss in field intensity due to out-of-plane radiation per unit of propagation distance. For this, we compute the loss in units of decibel per millimeter (mm) as 
\begin{equation}
L \left[\mathrm{\frac{dB}{mm}}\right] = \frac{10}{\ln 10} \left(\frac{\omega_k}{c}\frac{n_g}{Q_k}\right) \times 1\mathrm{mm},
\end{equation}
where $Q_k$ is taken from Fig. \ref{fig:daisy}(c). This quantity is shown in brown in panel (f) both for propagation along $\Gamma$M and $\Gamma$K. Results analogous to panels (a)-(e) but for the $\Gamma$K direction are shown in the Supplementary Information. We stress that, because the modes are BICs at $\Gamma$, the losses go to zero at the zero-index frequency. Furthermore, it is noteworthy that the loss is smaller than 10 dB/mm -- i.e. two orders of magnitude smaller than the propagation loss experimentally reported in Refs. \cite{Li2015, Vulis2017} -- over a wavelength bandwidth exceeding $30$nm, within which the effective index is $|n_e| < 0.1$. This shows that our structure is not only expected to have outstanding properties at the Dirac-cone frequency, but it could also be relevant for broadband applications.

\begin{figure}
\centering
\includegraphics[width = \columnwidth, trim = 0in 0in 0in 0in, clip = true]{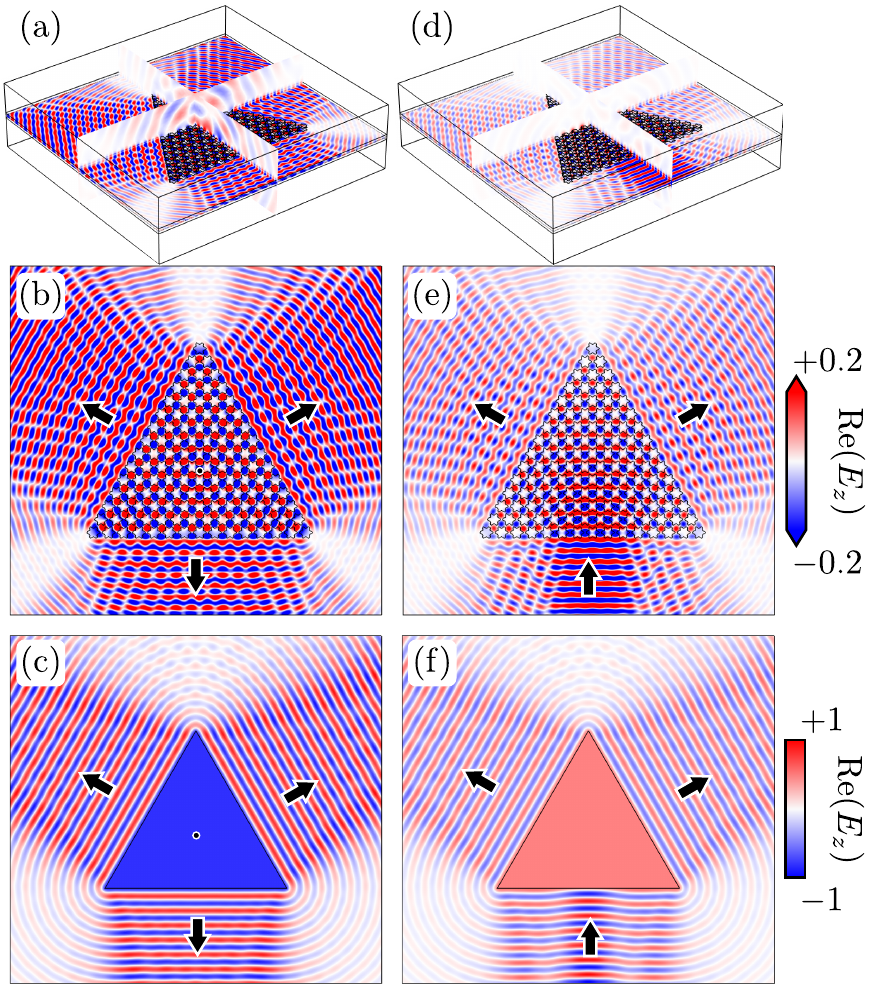}%
 \caption{Electric field Re$(E_z)$ for a simulated triangular region of zero-index PhC slab surrounded by un-patterned slab. (a): 3D view and (b): cross-section in the center of the slab for a dipole source excitation in the center of the PhC region (black dot), at frequency $\bar{\omega} = 0.486$. Arrows indicate the direction of the emitted waves. (c): Dipole-source excitation in the center of a homogeneous triangular region with relative permittivity and permeability $\varepsilon = \mu = 10^{-6}$. The simulated structure is infinite in the $z$-direction. (d)-(f): Same as (a)-(c), but with a Gaussian wavepacket excitation from below.} 
\label{fig:triangle}
\end{figure}

As a final set of demonstrations, we study the zero-index effect in a photnoic crystal structure that is finite in all in-plane dimensions, and compare these effects to those of a homogeneous zero-index material. The plots in Fig. \ref{fig:triangle}(a)-(b) show the emission due to an electric dipole source placed in the middle of a triangular region of the PhC, surrounded by un-patterned silicon slab. The dipole frequency is at the Dirac point, $\bar{\omega} = 0.486$. As expected for zero-index propagation, all of the elementary cells of the PhC oscillate in-phase, which leads to highly directional, collimated emission from the interfaces. This can be compared to panel (c), where we show a 2D simulation of a homogeneous triangular region with relative parameters $\varepsilon = \mu = 10^{-6}$ surrounded by air. Panels (b) and (c) match very well qualitatively. One difference worth noting is the fact that a near-field pattern is imprinted on the out-coming waves in (b), but this pattern gradually disappears, resulting in a plane-wave wavefront away from the interfaces. We also note that in panel (a), there is strong emission outside of the slab, but this is due to the \textit{direct} coupling of the source to radiative modes, as opposed to leakage of the Dirac-cone modes. To illustrate this point better, in panels (d)-(e) we show the same structure, but now illuminated by a Gaussian beam in the slab waveguide on the lower side. The characteristics of the outgoing waves in the uniform slab regions are similar to that of panel (b), and match well the simulation of a homogeneous region shown in panel (f). The important difference to note here is that in this case there is very little radiation outside of the slab. Most radiation outside the slab occurs around the in-coupling interface, which arises from a small impedance mismatch between the two regions. In this case, the absence of radiation on top of the interior of the photonic crystal slab indicates that once the zero-index PhC mode is excited, it has negligible radiation losses to the environment, as expected from Figs. \ref{fig:daisy}(e) and \ref{fig:ind_loss}(f).

In conclusion, we have shown a way to completely eliminate the out-of-plane radiation for an experimentally accessible zero-index structure. Combined with the fact that this is an all-dielectric device, our result is a significant step in the direction of recent efforts to introduce zero-index materials in integrated photonics platforms \cite{Li2015, Vulis2017}. Indeed, the large radiative losses have recently been identified as one of the main remaining challenges in that direction \cite{Vulis2018}. In our structure, the theoretically predicted losses go to zero at the Dirac-cone frequency, and, furthermore, for wavelengths in the $1550$nm range, we estimate propagation loss as low as $1$dB/mm and an index below $0.1$ within a $10$nm bandwidth. The presented photonic crystal is thus ideal  for applications in linear and nonlinear optics and emission control at optical and near-infrared frequencies.

\begin{acknowledgements}
This work was supported by the Swiss National Science Foundation through Project N\textsuperscript{\underline{o}} P300P2\_177721, by a Vannevar Bush Faculty Fellowship (Grant No. N00014-17-1-3030) from the U. S. Department of Defense, and by a MURI grant from the U. S. Air Force Office of Scientific Research (Grant No. FA9550-17-1-0002).
\end{acknowledgements}

%


\end{document}